\documentclass[10pt]{article}
\usepackage[english]{babel}
\begin{document}
\vskip9pt
{\Large \bf Anisotropic Vortices in High-Temperature
Superconductors and the Onset of Vortex-like Excitations above
the Critical Temperature}
\begin{center}
\vskip9pt
{\Large \bf Ondrej Hudak\footnote{\small e-mail: hudako@mail.pvt.sk} \\
Department of Theoretical Physics, Faculty of Mathematics and
Physics, Comenius University, Mlynska dolina F2, SK - 845 01
Bratislava}
{\Large \bf
\vskip2cm
Matej Hudak
\vskip1cm
Stierova 23, SK - 04023 Kosice
}
\end{center}

\vskip24pt
\newpage
\section*{Abstract}
In our recent papers we have found that a three-dimensional
superconducting state with anisotropic vortices localized
at the vortex-lattice points is a stable state
in zero and nonzero external magnetic field for the layered high-temperature
superconductivity materials.
There exists a phase transition at the temperature
$T_{c}^{v}$ from the normal phase to
the vortex superconducting state which is of the first order.
Note that the transition is in zero magnetic field.
The first order phase transition shows
overheating and overcooling effects. Nucleation of the
superconducting phase in the normal phase thus may occur at temperatures
higher than the transition temperature $T_{c}^{v}$. Then
the onset of the vortex-like excitations above
the transition temperature $T_{c}^{v}$ occurs in our theory.
The onset of the vortex-like excitations in Nerst signal and
some other experimental evidence for these excitations
above the transition temperature $T_{c}^{v}$
in LSCO, YBCO and in other
similar high-temperature materials may be explained thus by our theory.
The vortex-like excitations above and below
the transition temperature $T_{c}^{v}$ in LSCO, YBCO and in other
similar high-temperature materials continuously evolve.
This fact may be explained within our theory.

\newpage
\section{Introduction}

Sine-Gordon vortices may exist in superconductors
\cite{OHR} and \cite{OHRS}. These vortices may be present in
superconducting states also in zero magnetic fields. Their
topological charge is preserved. We studied a magnetic-field-induced
superconductive state in heavy fermion systems \cite{OH1} and
\cite{OH2}. The strength of the coupling between the order
parameter and magnetic field drives a superconductive state
induced by a magnetic field, \cite{OH1} and \cite{OH2}. Thus
such a state may be induced increasing the magnetic field.
Experimentally  it was reported that in $ CePb_{3}$
there exists a ferromagnetic phase and simultaneously there is
an evidence for presence of superconductivity. We studied
the Kondo-lattice superconductivity  in this connection,
\cite{OH2} and \cite{OH3}. Superconducting
states of different symmetry s, p and d were found to exist in the
Kondo-lattice systems. A phase transition from the normal phase to
the superconducting phase in $CePb_{3}$ type materials occurs in our
phenomenological model through creation of a vortex lattice, in
which vortices are superconducting regions induced by an
external magnetic field. Increasing the magnetic field the density
of such vortices and their shape becomes larger and the superconducting
phase increases its volume in the material.
In high-temperature superconductors, which are layered materials,
the interlayer distance is larger than the in-layer lattice
constant. The interlayer distance is comparable with the in-layer
penetration length. Our results  \cite{OH1} - \cite{OH3}
lead to study of formation of a vortex in the plane,
and to study of interaction of such
vortices in neighbouring layers. In the presence of an in-layer
vortex in-plane and out-of-plane magnetic field structures exist
due to currents which are present.
The vortex state which we described for a plane superconductor in
1982 may exist in zero magnetic field
\cite{OH}. This fact is very important difference from usual vortices
studied in superconductors, which are present due to an external
magnetic field. The stability of the vortex in zero magnetic field
\cite{OH} is due to the topological charge of this vortex which
is preserved. The magnetic field generated by this stable vortex
is influencing neighbouring planes and the plane in which it is
localized . This field is coupled with the superconducting order
parameter due to symmetry. As we discussed above the strength
of the coupling between the magnetic field and the order
parameter may drive a superconductive state with
vortices induced by the magnetic field of this vortex, \cite{OH1} and \cite{OH2}.
Thus in layered high-temperature superconductors coupling between
the magnetic field generated by this stable vortex and the order
parameter may lead to generation of other vortices
(antivortices). This is a selfconsistent process.
The field of these vortices influences this stable vortex.
Thus then we may study vortices and their in-plane and
out-of-plane interactions due to the inductance coupling.
Thus the strength of the coupling between the order
parameter and magnetic field drives a superconductive state
induced by a magnetic field as mentioned above, \cite{OH1} and
\cite{OH2}, and in layered high-temperature superconductors
drives a superconducting state with vortices.
Note that also in the zero
external magnetic field such a selfconsistent process may exists.
Creation of a vortex in one plane leads to magnetic fields which
influence the neighbouring planes, in these neighbouring planes
these fields may induce creation of vortices the magnetic fields of
which influence the original vortex in the plane: its existence and
properties.
Creation of a vortex in the plane leads also to magnetic fields which
influence the same plane, in this plane
these fields again may induce creation of vortices
the magnetic fields of which influence the original vortex in the
same plane: again its existence and properties.

Thus it follows that in layered materials one may expect
that formation of such vortices may stabilize   the superconducting state
of different from classical  symmetry in zero external magnetic field.
Note that in nonzero external magnetic fields the self-consistent
process may be modified but still present.

The vortex in the superconductor plane which we studied
\cite{OH} is anisotropic vortex. This anisotropy (symmetry
different from symmetry of isotropic vortices)
then leads to coupling to other quantities of the system for
which an invariant contributing to free energy expansion may
exist. Integrating out these other quantities from the free energy
of the superconducting material we obtain the effective free energy
expansion for the superconducting order parameter with anisotropy
terms present.

Recently we described and studied the mentioned mechanism and
formulated a general problem: to study behaviour of such
anisotropic vortices in zero and non-zero magnetic fields.
The vortices of the in- and inter-
layer type should be compared as concerning their theoretical
properties in zero and non-zero fields with existing experiments
concerning study of vortices in high-temperature superconductors,
 \cite{KAM}, \cite{JRK}, \cite{AAT} and \cite{KA1}.
It was found experimentally  that the anisotropy of the
interplane-vortices in these
materials is intrinsic. This is natural, the vortex is parallel
to layers, layers are anisotropic. Planes are involved in such
vortex configuration, we are speaking about the inter-plane vortex induced by
an external magnetic field. As far as we know the anisotropy of
such vortices is interpreted as due to anisotropy of layers.

The vortices which we were studied theoretically
are intrinsically  anisotropic. However this anisotropy is
present for vortices which are present within an isotropic plane.
Their anisotropy is not generated due to anisotropy of the material.
Thus it is important to compare experimental
results concerning anisotropy properties of the vortices in the layered
materials with those properties of vortices which we found
theoretically. This is the reason why we reviewe
different superconducting states with anisotropic vortices:
the superconducting states in which single vortex
exists, the superconducting states in which lines and lattices of
vortices exist, the superconducting states in which pairs of
vortex-antivortex type are present and form a kind of liquid.
At larger distances between vortex-antivortex
in the pair the energy is linearly dependent on this distance.
This is a kind of confinement phenomena known from quarks forming
baryons.  For smaller distances there exists a
minimum distance for which the vortex-antivortex pair exist in a
ground state. The pair vortex-antivortex thus does not
annihilate. Further we considered in our papers
the superconducting states in which solitons and
strings of vortices, which represent
some kind of structures induced by internal mechanisms /twinning of
the crystals, etc./ or by surface, may exist.

Intrinsic Josephson effect in layered systems was described in
\cite{LD} and a summary of the Josephson effect in the
high-temperature superconductors is given in \cite{LNB}.
The time dependent equations for the gauge invariant phase differences
of the layered order parameters for superconductive states and for a
general vector potential of the electromagnetic field we discussed
in \cite{OHR} and \cite{OHRS}. The interlayer coupling within
the mentioned model
equations was studied with its influence on the properties of vortices in
layers. These vortices in neighbouring layers are coupled. They
are forming a string of vortices perpendicular to the layers in
which the vortices exist. The string goes through the centers of
these vortices.

We
used in our study a small parameter expansion method, and an exact method
of solution of the corresponding Lagrange-Euler equations to compare different
macroscopic superconducting states. We have found their free energy
and discussed their properties. The corresponding
Lagrange-Euler equations are nonlinear coupled sine-Gordon equations
in the approximation of the constant amplitude of the order
parameter.
Let as note that the nonlinearity describing the vortex
configurations in nonzero /and in zero/ magnetic field is
neglected due to difficulties of the mathematical origin and mainly due
to expectation that the nonlinearity is negligible around
the core of the vortex, which is assumed to be large.
However we have found that nonlinearities in order parameter interactions
are physically essential, \cite{OH}. The Lagrange-Euler
equations for the superconducting state order parameter and for
the vector potential of the electromagnetic field
in the constant amplitude approximation for the
scalar (s - type) superconducting order parameter reduced in the
model studied in our papers cited above to equations which are
similar to the two-dimensional sine-Gordon equation. They are
generalized coupled two-dimensional sine-Gordon equation.
p and d symmetric superconducting states were not studied until now.

The two-dimensional sine-Gordon
equations occur in other physical systems,
in quantum antiferromagnets, in classical
two-dimensional XY model with the in-plane magnetic field \cite{MBG} -
\cite{PP}, in models describing crystal growth \cite{OH4}
and \cite{DK}, and in models describing defects in incommensurate
systems \cite{BK} and \cite{BK1}. It is useful to use analogy and
mathematical properties of solutions of these equations for study
of anisotropic vortices in superconductors.

In our paper from 1982 we used a new method for solving a two-dimensional
sine-Gordon equation  \cite{OH}. We transformedg this equation into a
set of two second order ordinary nonlinear non-coupled equations.
Multi - (Resonant-Soliton) solutions and vortex-like solutions
of the two- and three - dimensional sine-Gordon
equations were studied in \cite{T}. The structure
of a vortex in terms of the solitons was studied in \cite{N}.
The vortex is formed from solitons which are intersecting.
The multi-vortex solutions of the sine-Gordon equation
were studied in \cite{BT}.
Exact rational-exponential solutions of this equation
were studied in \cite{B}. Vortices
and the boundary value problem were studied by the inverse scattering
transform method for this type of equation in \cite{BTC}. The
singular solutions of the elliptic sine-Gordon equation were studied
numerically  in \cite{AGS}. Numerical studies of dynamical
isoperimeter pattern are in \cite{TE}.

Vortex-like excitations in $ La_{2-x}Sr_{x}CuO_{4} $ at
temperatures significantly above the critical temperature were
found in the Nerst effect signal \cite{XOWKU}. In overdoped
$ La_{2-x}Sr_{x}CuO_{4} $ (LSCO) the upper critical field $ H_{c2} $
does not end at $ T_{c0} $ but at a much higher temperature.
These results imply, according to \cite{WOXKUBLH}, to a loss in
phase rigidity rather than in a vanishing of the pairing amplitude.
Nerst measurements in $ YBa_{2}Cu_{3}O_{4} $ (YBCO) and in
$ Bi_{2}Sr_{2-y}La_{y}CuO_{6} $, \cite{WXKUOAO},
and in LSCO in fields up to 33 T show the existence
of vortex-like excitations high above $ T_{c0} $.
This anomaly is related to the key question
of whether the Meissner transition in
zero field is caused by the collapse of long-range phase
coherence or by the vanishing of the pairing amplitude.
In zero magnetic field there is equal number of
vortices with plus and minus topological charge. In nonzero
fields there is nonequal number of vortices with plus and minus
topological charge.

Preformed pairs as superconductor fluctuations may exist in Bose
condensation of localized  Cooper pairs with short coherence
length, \cite{YJU}, \cite{ASA} and \cite{MR}. Pairing
correlations without phase coherence are described in \cite{GB}
and \cite{EK}.

In our papers \cite{OHR} and \cite{OHRS}.
new results were found as concerning the
influence of neighbouring planes on the vortex state in a given
plane and new exact solutions of the coupled Lagrange-Euler
sine-Gordon equations were found, their properties and free energy
were compared. We have found that the three-dimensional
superconducting state with vortex lattice, which is a stable
state in zero external magnetic field for the layered materials.
There exists a phase transition at the temperature
$T_{c}^{v}$ from the normal phase to
the vortex superconducting state which is of the first order.
Depending on the parameters of the material the transition
temperature $T_{c}^{v}$ may be higher than zhat in the case of the
second order phase transition from normal phase to the
homogeneous superconducting phase without vortices.
The transition temperature $T_{c}^{v}$ would be zero for zero
topological charge of vortices. Note that the transition to
vortex lattice state is present also in
zero magnetic field. The first order phase transition shows in
experiments usually overheating and overcooling effects. Nucleation of the
superconducting phase in the normal phase occurs at temperatures
higher than the transition temperature $T_{c}^{v}$. And
vice-versa. Nucleation of the
normal phase in the superconducting phase occurs at temperatures
lower than the transition temperature $T_{c}^{v}$.
Thus the onset of the vortex-like excitations above
the transition temperature $T_{c}^{v}$ occurs in our theory due
to nucleation of the superconducting phase with vortex lattice
(or parts of this lattice) in the normal phase
at temperatures higher than the transition temperature $T_{c}^{v}$.
The onset of the vortex-like Nerst signal above
the transition temperature $T_{c}^{v}$ in LSCO, YBCO and in other
similar high-temperature materials may be thus explained by our theory.

In this paper we review our results on
vortex configurations in two-dimensional sine-Gordon
systems in connection with high-temperature superconductors.
We will use this results in our discussion concerning anisotropic
vortices and the onset of Nernst signal and of some other signals
corresponding to presence of vortex-like excitations above the
critical temperature. The superconducting planes are coupled
by the Josephson effect. The Lawrence-Doniach functional leads to
the Lagrange-Euler equations for the order parameter.
In \cite{OHR} and \cite{OHRS}.
We discussed the vortex states using exact solutions
of these coupled Lagrange-Euler sine-Gordon equations.
Properties of the exact solutions and the free energy of the
corresponding order parameter configurations weres compared with the
normal phase state and the homogeneous superconducting state.
Stable vortex configurations in zero external magnetic field for
the layered materials exist,  the same holds for a lattice of vortices
case.
The first order phase transition overheating and overcooling
effects are present here. Nucleation of the
superconducting phase in the normal phase, which occurs at temperatures
higher than the transition temperature $T_{c}^{v}$, exists.
The onset of the vortex-like excitations above
the transition temperature $T_{c}^{v}$, which occurs in our
theory, is discussed. The onset of the vortex-like Nerst signal above
the transition temperature $T_{c}^{v}$ in LSCO, YBCO and in other
similar high-temperature materials is not well understood in
experiments. The vortex-like excitations above
the transition temperature $T_{c}^{v}$ in LSCO, YBCO and in other
similar high-temperature materials continuously evolve into
vortex states below
the transition temperature $T_{c}^{v}$, this fact may be
explained within our theory. Our theory may explain this fact and
some of other experimentally observed properties of high-temperature
superconductors described above.

\section{Superconducting Planes Coupled by the Josephson
Effect}

Superconducting layers in layered superconductors
of the high-temperature type are
coupled due to intrinsic Josephson effect. The Lawrence - Doniach
free energy functional for the order parameter for such a system has the form
 \cite{OHR}

\begin{equation}
\label{1} F (\Psi_{n}(\bf r), {\bf A}({\bf R}) ) =
\frac{H^{2}_{c}s}{4 \pi} \sum_{n} \int d{\bf r} [ \zeta^{2}_{ab}
\mid (-i \nabla + \frac{2 \pi}{\Phi_{0}}{\bf A}_{n}) \Psi_{n})
\mid^{2}-
\end{equation}
\[ - \mid  \Psi_{n}) \mid^{2} + \frac{1}{2}\mid  \Psi_{n}) \mid^{4} + \]
\[ + r (\mid  \Psi_{n}) \mid^{2} + \mid  \Psi_{n+1}) \mid^{2} - \Psi_{n} \Psi^{*}_{n+1} \exp(-i \chi_{n,n+1}) - \Psi^{*}_{n} \Psi_{n+1} \exp(+i \chi_{n,n+1}) ) \]
\[ + \int d{\bf R} \frac{B^{2}}{8 \pi} \]

The order parameter in the n-th layer is $ \Psi_{n}({\bf r}) =
\mid \Psi_{n}({\bf r}) \mid \exp(i \Phi_{n}](\bf r))$, here z=ns,
and s is an interlayer distance, {\bf R} = ({\bf r}, z), $ \nabla =
\frac{\partial  }{\partial {\bf r}}$, $ H_{c} $ is the bulk critical
field, $ \zeta_{ab} $ is the coherence length in the ab plane which
is perpendicular to the c-direction. $ \Phi_{0} = \frac{\pi \hbar
c}{ \mid e \mid} $, $ {A}_{n} (\bf r) = (A_{nx}, A_{ny})$ is an
average value of the vector potential over the distance
$ [(n - \frac{1}{2})s, (n +  \frac{1}{2})s $ along the c-axis, {\bf
B} = rot~({\bf A}), and $ \chi_{n,n+1} = \frac{2 \pi}{ \Phi_{0}}
\int^{(n+1)s}_{ns} dz A_{z} $.

If we denote the Josephson coupling constant
between neighbouring layers order parameters as r, where $ r<<1 $,,
the coherence length $ \zeta_{c} $ in the c-direction is given by

\begin{equation}
\label{2}
\zeta^{2}_{c} = \frac{s^{2}r}{2}
\end{equation}

The in-plane penetration depth $ \lambda_{ab} $ is given by

\begin{equation}
\label{3}
\lambda^{2}_{ab} = \frac{\Phi^{2}_{0}}{8\pi^{2} H^{2}_{c}
\zeta^{2}_{ab}}
\end{equation}

The anisotropy ratio $ \frac{\lambda_{ab}}{\lambda_{c}} $ is given by

\begin{equation}
\label{4}
\frac{\lambda_{ab}}{\lambda_{c}} = \frac{\zeta_{ab}}{
\zeta_{c}}
\end{equation}

The interlayer coupling is weak, $\zeta_{c} << s $. The general
case and the limit of strong interlayer coupling
$s <<  \zeta_{c} $ will not be discussed here.

 Let $ \phi_{n}(\bf r) $ be the change of the order parameter
phase around a vortex. The constant amplitude approximation $ \mid
\Psi \mid^{2} = 1$ we consider first. The
constant amplitude approximation in which the density of
superconducting pairs is temperature dependent will be considered
at the end of the paper. There is
experimental evidence that the constant amplitude approximation
is quite a good approximation to understand  vortex state
(phase), see discussion below.

\section{Lagrange-Euler Equations for the Order Parameter}

The Lagrange-Euler equations for the phase difference $
\phi_{m,m+1}$ are the sine-Gordon coupled equations

\begin{equation}
\label{8}
- \sum_{m} L_{n,m} \nabla^{2} \phi_{m,m+1} +
\frac{1}{\lambda^{2}_{J}} \sin(\phi_{n,n+1}) = 0
\end{equation}

where the summ over m is over layers.

The interlayer inductance $ L_{n,m} $ between the layers n and m
has the form

\begin{equation}
\label{9} L_{n,m} = \int^{2 \pi}_{0} \frac{dq}{2 \pi}
\frac{\cos(n-m)q}{2(1-cos(q))+\frac{s^{2}}{\lambda^{2}_{ab}}} =
\frac{\lambda_{ab}}{s} (1 - \frac{s}{\lambda_{ab}})^{\mid n-m \mid}
\end{equation}

The boundary conditions for the Lagrange - Euler equations (\ref{8})
are

\begin{equation}
\label{10} (\nabla_{x} \nabla_{y} - \nabla_{y} \nabla_{x})
\phi_{n,n+1}(\bf r) = 2 \pi \sum_{n} ( \mid \delta( {\bf r} - {\bf
r}_{n \nu})  - \delta( {\bf r} - {\bf r}_{n \nu}) \mid )
\end{equation}

The equations (\ref{8}) are thus a system of
the sine-Gordon equations in two dimensions (planes ab)
coupled in the perpendicular direction c.

\section{The Distance s is Comparable with the Penetration
Depth $ \lambda_{ab} $}

When the distance s is comparable with the penetration depth $
\lambda_{ab} $ as it is the case in high-temperature
superconductors the mutual interlayer inductance $ L_{n,n} $
(which is in this case when m=n the inlayer inductance in fact)
is nonzero and much larger than the other inductances
(interlayer, because m and n are different) $ L_{n,m} $

\begin{equation}
\label{11} L_{n,n} =  \frac{\lambda_{ab}}{s}
\end{equation}

Thus inductances  $ L_{n,m} $ with n different from m are
not taken into account here in this section.

For the phase difference $ \phi_{m,m+1} $ the equations have in this
case the form

\begin{equation}
\label{12}
 -  L_{n,n} \nabla^{2} \phi_{n,n+1} +
\frac{1}{\lambda^{2}_{J}} \sin(\phi_{n,n+1}) = 0
\end{equation}

where the lower order terms in $\frac{L_{n,m}}{L_{n,n}}$ for n
and m different are neglected.
They would lead to multi-Sine-Gordon equations.

Let us denote by

\begin{equation}
\label{13} \phi = \phi_{n,n+1}
\end{equation}

an angle variable. We assume that this angle is the same for every
plane, which may be done due to expected homogeneity in the sense
of the same states present in all layers.

Let us denote further

\begin{equation}
\label{14} \lambda_{p} = \lambda_{J}.L_{n,n}
\end{equation}

The two-dimensional sine-Gordon equation (\ref{12}) takes the form

\begin{equation}
\label{15} \triangle  \phi =  \frac{1}{\lambda^{2}_{p}} \sin(\phi)
\end{equation}

This form of the two-dimensional sine-Gordon equation was
studied by the author in \cite{OH}.

\section{The Distance s is Comparable with the Penetration
Depth $ \lambda_{ab} $ - Influence of Next Layers}

In the previous section we described lagrange-Euler equations
for the case when the distance s is comparable with the penetration depth $
\lambda_{ab} $. We neglected the influence of neighbouring layers.
Let us consider now the effect of nearest neighbour layers.
The interlayer inductance for neighbouring layers in this case
$ L_{n,n \pm 1} $ is

\begin{equation}
\label{11'} L_{n,n \pm 1} =  \frac{\lambda_{ab}}{s}(1-
\frac{s}{\lambda_{ab}} )
\end{equation}

The equation for the phase differences $ \phi_{m,m+1} $ has
the form

\begin{equation}
\label{12'} -  L_{n,n} \nabla^{2} \phi_{n,n+1} -  L_{n,n+1}
\nabla^{2} \phi_{n+1,n+2} -  L_{n-1,n} \nabla^{2} \phi_{n-1,n}+
\frac{1}{\lambda^{2}_{J}} \sin(\phi_{n,n+1}) = 0
\end{equation}

Let us denote

\begin{equation}
\label{14'} \lambda_{p} = \lambda_{J}.\sqrt{L_{n,n}}
\end{equation}

The two-dimensional sine-Gordon equation (\ref{12'}) takes now
the form

\begin{equation}
\label{15'} -  \nabla^{2} \phi_{n,n+1} -  l \nabla^{2}
\phi_{n+1,n+2} -  l \nabla^{2} \phi_{n-1,n}+
\frac{1}{\lambda^{2}_{p}} \sin(\phi_{n,n+1}) = 0
\end{equation}

where the parameter l is small,  and it is defined as

\begin{equation}
\label{11''} l =  (1- \frac{s}{\lambda_{ab}} )^{1}
\end{equation}

We left the exponent 1 in order to indicate how the next to the
nearest layers will influence the Lagrange-Euler equations.
In this form of the two-dimensional sine-Gordon equation three
neighbouring planes are coupled explicitly. The solution
of the equation (\ref{15'}) is again looking in the homogeneous
 form, in which we again introduce the variable

\begin{equation}
\label{13'}
\phi = \phi_{n,n+1}
\end{equation}

for every n. This again may be done due to expected homogeneity in
the sense mentioned above as concerning the states in planes
in this limit.

The equation (\ref{15'}) takes now the form

\begin{equation}
\label{15''} -  \nabla^{2} \phi + \frac{1}{\lambda^{'2}_{p}}
\sin(\phi) = 0
\end{equation}

where we defined the renormalized Josephson length

\begin{equation}
\label{14''} \lambda^{'}_{p} = \lambda_{J}\sqrt{1+2l}
\end{equation}

which is larger than $\lambda_{p}$ due to the fact  that l is
positive. Note that the larger l the larger renormalised Josephson
length.

This homogeneous solution considered in this section
has the characteristic length $\lambda^{'}_{p}$
larger than in the one-layer case length $\lambda_{p}$ due
to the inter-layer coupling via interlayer inductance.

The free energy of the single vortex in every layer and for
the same homogeneous state (homogeneous as concerning dependence
on the the number n of the layer, the vortex state is off course
nonhomogeneous state) in all layers depends linearly on L, the system
linear dimension

\begin{equation}
\label{14.1} 2^{\frac{3}{2}}N \pi J 2^{2} \frac{L}{\lambda^{'}_{p}}-
2k_{B}T N\ln(\frac{L}{a_{0}})
\end{equation}

where N is the number of layers, the cutoff constant $a_{0}$ is a
constant of the order of lattice constant.

We see that if the coupling between layers is taken into account
 then the free energy is lower due to smaller energy of the vortex.
The interlayer coupling decreases the free energy and stabilizes the
state with vortices the cores of which are localizes on the same
line in the direction c perpendicular to the planes ab. Taking into
account even higher order inductance couplings it is expected that further
lowering of the free energy will be found and that this will
further stabilize  the vortex state.

Let us consider the solution of the equation (\ref{15'}) in which
the sign of the angle $\phi$ changes in neighbouring layers.
This solution describes a nonhomogeneous sate in the sense that
the order parameter angle changes its sign in neighbouring planes
, e.i. it is shifted by the angle $\pi$

\begin{equation}
\label{13''} \phi_{n,n+1}= (-1)^{n}\phi
\end{equation}

for every n. The equation (\ref{15'}) takes in this case the form

\begin{equation}
\label{15'''} -  \nabla^{2} \phi + \frac{1}{\lambda^{''2}_{p}}
\sin(\phi) = 0
\end{equation}

where the Josephson characteristic length has now the form

\begin{equation}
\label{14'''} \lambda^{''}_{p} = \lambda_{J}\sqrt{1-2l}
\end{equation}

This Josephson length is smaller than the Josephson length
$\lambda_{p}$, we still assume l to be a small parameter.
Thus the solution in which the sign of angle $\phi$ changes
in neighbouring layers has the length $\lambda^{''}_{p}$ smaller
than the length $\lambda^{'}_{p}$ and
smaller than the original in-plane Josephson length $\lambda_{p}$.
The free energy of the state with a single vortex in a given layer
and with minus sign in neighbouring layers depends again linearly on L, the
system linear dimension

\begin{equation}
\label{14.2} 2^{\frac{3}{2}}N \pi J 2^{2}
\frac{L}{\lambda^{''}_{p}}- 2k_{B}T N\ln(\frac{L}{a_{0}})
\end{equation}

where again N is the number of layers, the cutoff constant $a_{0}$
is a constant of the order of the lattice constant.

If the inductance coupling l between layers is taken
into account then we have found that the free energy of the state in which
 the sign of the angle $\phi$ changes in neighbouring layers is higher
with respect to the state in which
 the sign of the angle $\phi$ does not change in neighbouring layers.
The free energy of the state in which
 the sign of the angle $\phi$ changes in neighbouring layers is also higher
with respect to the state in which the layers are not coupled at
all.

The small parameter l may be taken as an expansion parameter when
solving the equation (\ref{15'}). One finds in that case that the
homogeneous in above sense state with a vortex in one layer and
with other vortices in other layers with their core center
in the same point in the plane ab /a line
of vortices in the direction c/ is more stable.
Single-vortex state in one layer and
the same state in all layers give lower free energy due to the
interlayer inductance coupling. The interlayer inductance
coupling plays stabilisation role for these vortices.
One can ask now that if many-vortex states in planes
are realized, which distance between vortices is such that
the vortex superconducting state becomes more preferable than the
normal state, or than the homogeneous superconducting state without vortices
(usual superconducting state in classical superconductors). The
free energy of the single vortex in a given layer and the same
(homogeneous case as concerning the signs of the phase variable)
state with a vortex in neighbouring layers depends
linearly on L, the system linear dimension and the (critical) value of
the parameter l. At a given temperature it is given by the condition
that the free energy with the vortex at the same place in every
layer is the same as the free energy of the superconducting state
without vortex

\begin{equation}
\label{14.3} F(vortex)=2^{\frac{3}{2}}N \pi J q^{2}
\frac{L}{\lambda^{''}_{p}}- 2k_{B}T N\ln(\frac{L}{a_{0}})=F(hom)=0
\end{equation}

For a given temperature and other parameters of the system
increasing the parameter l decreases the energy in the free energy,
it may happen then that the vortex state becomes the most stable state.
It is clear that the transition to the superconducting state
occurs at higher temperature than for the state without vortex. The
transition to the single-vortex state with the inter-layer coupling
occurs at higher temperature than for the state in which the
single-vortex superconducting state does not take into account the
inter-layer inductance coupling.

If the density of superconducting pairs depends on temperature, we
introduce $\alpha = \alpha_{0}(T-T_{c})$, $T_{c}$ is the bulk
transition temperature, $\beta$ is a constant, the free energy
$F(hom)$ is found from (\ref{1}) for $\mid \Psi \mid^{2} = \frac{\mid \alpha
\mid}{\beta}$ . The Lawrence-Doniach functional for the order parameter
with $\Phi$ constant gives

\begin{equation}
\label{5''} F (\Phi_{hom} ) = - \frac{H^{2}_{c}sL^{2}N \alpha^{2}}{8
\pi \beta}
\end{equation}

where $H_{c} $ is the bulk critical magnetic field.
The constant J and the bulk critical magnetic field value
$H_{c}$ are related

\begin{equation}
\label{14.4} J= \frac{H_{c}^{2}s \zeta^{2}_{ab}\alpha^{2}}{4 \pi
\beta}
\end{equation}

Here we consider the constant amplitude approximation $\mid \Psi
\mid^{2} = \frac{\mid \alpha \mid}{\beta}$ instead of $\mid \Psi
\mid^{2} = 1$, we assume that temperature is near the transition temperature.
The gradient term in the phase in this case gives also contribution
to the free energy. It is of the order of $\frac{4s}{\lambda}$ as a
contribution to the bulk critical temperature $T_{c}$.
Note that the case with non-zero magnetic field may be
easily described and will not be considered here.

The normal state is given by $\Psi=0$ and its free energy is 0. The
superconducting state with the homogeneous (in the classical
sense) superconducting order parameter without vortices
is given by $\mid \Psi \mid^{2}=
\frac{\mid \alpha \mid}{\beta}$ and its free energy is

\begin{equation}
\label{14.6} F(hom)=-\frac{NsL^{2}\zeta^{2}_{ab}H^{2}_{c}}{8\pi}=
-\frac{JNL^{2}\alpha^{2}}{2 \beta \zeta^{2}_{ab}}
\end{equation}

The vortex state with the $N_{v}$ vortices cores of which are
localized in the square /for simplicity/ lattice
has the free energy

\begin{equation}
\label{14.7} F(N_{v}-vortex)=N_{v}N[2^{\frac{3}{2}} \pi J q^{2}
\frac{L_{v}}{\lambda^{'}_{p}}- 2k_{B}T \ln(\frac{L_{v}}{a_{0}})]
\end{equation}

The inter-vortex distance $L_{v}=\frac{L}{N_{v}}$ can be found
minimizing the free energy $F(N_{v}-vortex)$ with respect
to $L_{v}$. We have found the
minimal value $L_{v}$

\begin{equation}
\label{14.8} L_{v}= \frac{\sqrt{2}\lambda^{'}_{p}k_{B}T}{\pi J
q^{2}} ln(\frac{\sqrt{2}\lambda^{'}_{p}k_{B}T}{\pi J q^{2}a_{0}})
\end{equation}

The inter-vortex distance /vortex lattice constant/ is increasing
with temperature T and is decreasing with temperature T depending whether
temperature is above or below the critical temperature
$T^{v}_{c}$, respectively. The critical temperature $T^{v}_{c}$
is given by

\begin{equation}
\label{14.9}
T^{v}_{c} = \frac{\pi J
q^{2}a_{0}}{\sqrt{2}\lambda^{'}_{p}k_{B}e}
\end{equation}

 The free energy for the vortex lattice state with $N_{v}$
 vortices in the plane  has the form

\begin{equation}
\label{14.10} F(N_{v}-vortex)=
NL^{2}[\frac{2\pi^{2}J^{2}q^{4}ln(\frac{\sqrt{2}\lambda^{'}_{p}k_{B}T}{\pi
J
q^{2}a_{0}})}{\lambda^{'2}_{p}k_{B}T}-\frac{\pi^{2}J^{2}q^{4}}{\lambda^{'2}_{p}k_{B}T
ln^{2}(\frac{\sqrt{2}\lambda^{'}_{p}k_{B}T}{\pi J q^{2}a_{0}})} ]
\end{equation}

The free energy of the $N_{v}$ - vortex state should be compared
 with the free energy $F(N)=0$ of the normal (metal) state,
 and with the free
energy of the homogeneous (in the classical sense)
superconducting state without vortices given by

\begin{equation}
\label{14.11} F(sc)= -\frac{JNL^{2}\alpha^{2}}{2 \beta
\zeta^{2}_{ab}}
\end{equation}

As can be seen the transition at the transition temperature $T_{cv}$ from the
normal phase to the $N_{v}$ - vortex superconducting state occurs,
it is of the first order.

Depending on the parameters of the material the transition temperature
$T_{cv}$ may be higher for this transition than the transition temperature
in the case of the second order phase transition from the normal phase to the
homogeneous superconducting phase without vortices. It is given
by

\begin{equation}
\label{14.12} T_{cv}= \frac{\pi J
q^{2}a_{0}}{\sqrt{2}\lambda^{'}_{p}k_{B}}
\end{equation}

Note that the transition temperature $T_{cv}$ would be equal zero
for zero topological charge q. Note further that the transition
temperature $T_{cv}$ increases when the interlayer inductance
increases.

We will now discuss several solutions of the two-dimensional
Sine-Gordon equations.

\section{Single-Vortex Configurations}
In \cite{OH} we have found the single vortex configuration as a
solution of the two-dimensional sine-Gordon equation, and thus
of the order parameter. It has the form

\begin{equation}
\label{16} \phi = \pm \tan^{-1}(\frac{\sqrt{\frac{a}{(2-a)}}
\sinh(\sqrt{\frac{2-a}{a}} \frac{(x-x_{0})}{\lambda})}
{\sinh(\sqrt{\frac{a}{2}} \frac{(y-y_{0})}{\lambda })})
\end{equation}

here $\lambda$ equals $\lambda_{p}$. The vorticity of this vortex is
plus minus four. In the limit
 of $ \lambda >> L $ we obtain a usual isotropic vortex configuration.
The configuration (\ref{16}) breaks the rotational symmetry.
The free energy of the configuration given by (\ref{16})
is linear in L. This is in difference to isotropic vortices.
In the limit of $ \lambda >> L $ where a usual
isotropic vortex appears the logarithmic dependence of the free
energy of the configuration on L is found. Thus in systems in
which anisotropic vortices are present and which are small, e.i.
in which $ \lambda >> L $, the anisotropy of the vortex will be
almost undetectable.

In the limit of $\lambda_{p} \gg L$ the anisotropic vortex takes
an isotropic form is the zeroth approximation.

\begin{equation}
\label{14.13} \phi = \tan^{-1}(\frac{(x-x_{0})}{(y-y_{0})})
\end{equation}

In isotropic material systems in which anisotropic vortices
are present and which are large enough, e.i. in which $ \lambda << L $,
the anisotropy of the vortex will be detectable.

\section{Multi-Vortex Configurations}

Multi - (Resonant-Soliton) - Soliton solutions and vortex-like
solutions in two and three dimension for the sine-Gordon equation
were studied in \cite{T}. For the two- and three- dimensional
sine-Gordon equations there exist exact multi - (resonant-soliton) -
soliton solutions and vortex-like solutions, in addition to exact
multi-soliton and resonant - soliton solutions.
In \cite{N} it was shown that the quasi-vortex type solutions from
\cite{OH} can be derived from the multiple soliton solutions by the
proper procedure. Thus it follows from that there exist multiple
vortex-like solutions. Moreover an anisotropic vortex
configuration far from the core of the vortex has the form of
almost unchanging order parameter regions separated by
soliton-like walls.

\section{Static Lattice of Vortices, Multi-vortex Solutions}

Let us briefly overview solutions of the sine-Gordon equations
which represent static lattice of vortices and multi-vortex
solutions. These solutions and  space periodic vortex chains and lattice
and breather solutions were studied in \cite{BT}. The boundary
problem for two-dimensional sine-Gordon equation was studied by the inverse
scattering method in \cite{ABB}.
Relation between vortex solutions of the two-dimensional sine-Gordon
equation and solitons of the same equations were studied in
\cite{N}. Single vortex solution \cite{OH} can be derived from the
known multiple soliton solutions by a proper procedure. This fact
shows possibility to find multi-vortex solutions of the
two-dimensional sine-Gordon equations due to existence of
multisoliton solutions.

Vortex-type lattice solutions found by Hirota method and by Backlund
transformation represent alternating vortices with $q= \pm 4$ and
form a tetragonal or square lattice. Interaction of the spin-wave
and vortex lattice are similar to the soliton-wave interactions.

Vortex-antivortex pair with the vorticity $q=\pm 4 $ is given by

\begin{equation}
\label{14.15} \phi = \tan^{-1}(\frac{1-x+\sinh(y)
\exp(-x)}{\sinh(y)-(1+x)\exp(-x)})
\end{equation}
The total topological charge of the vortex-antivortex pair is
zero. Such pairs may be in pre-vortex-lattice states, as a kind
of liquid.

A vortex array was found with topological charges $ q = \pm 4 $.
As far as the author is aware the vortex arrays were not observed
directly, however several vortex arrays which are parallel may
have an equilibrium distance between arrays and thus may be
difficult to distinguish them from the tetragonal phase at first
sight. However one may expect that stiffness of such
configuration in the direction of arrays
will be substantialy different from that in the perpendicular
direction.
Numerical singular solutions of the elliptic sine-Gordon equation
\cite{AGS} were studied. A semi-infinite $2\pi$ -
kink starting from a singular point of the vortex type was found.
Ring-like $2\pi$ - kinks with a common center in the singular point
were found. The singular point with a linear region of the
logarithmic spiral was found. Exact solutions of the sine-Gordon
equations as rational-exponential solutions are described in
\cite{GWB}. Exact rational-exponential solutions of the
two-dimensional sine-Gordon equations are constructed by a method
based on the formal perturbation method. These solutions represent
the nonlinear superposition of two interacting $2\pi$ kinks, at the
point of intersection we have a vortex with topological charge plus
and minus 4. Numerical analysis shows that the minimum distance
between centers of vortices with opposite topological charges in
dipole state exists and is given by approximately 1.1 in lambda
units. These two vortices cannot annihilate in spite of zero total
topological charge of the dipole.

\section{Dynamic Properties of Two-Dimensional Sine-Gordon
Vortices}

The two-dimensional sine-Gordon equation with time dependence was
studied in \cite{TE}, and it was found that a particular dynamical
pattern in a perturbed two -dimensional sine - Gordon systems
exists. The stability of vortex-like solutions were checked.
The main topological invariant of the vortex-like solutions is
due to the total length of the $ \pm 2 \pi $ wave fronts entering as elementary
kink-like patterns in the constitution of the whole
configuration. This corresponds to the energy conservation law.
In \cite{TE} it was also shown that a particular dynamical pattern in a
perturbed or not two-dimensional sine-Gordon system exists. Its
stability with respect to perturbations and in time was studied.
Their main topological invariant is the total length of the kink
wave fronts forming the pattern and the invariant is consistent with
the conservation of energy in all situations.

In \cite{T} the time dependent sine-Gordon equation without damping
was studied. The procedure like that used in \cite{N} leads to a
two-soliton solution, thus intersection of these solitons
corresponds with a moving vortex. On the other hand the author
\cite{T} shows that three or higher number soliton
solutions are known to exist
only with limited parameter range. Thus time dependence without
damping here does not lead to a vortex-lattice in general, this may exist
only in a limited range of parameters.

\section{Summary and Discussion}

In our papers \cite{OHR} and \cite{OHRS} we studied layered
superconducting materials. Vortex-like states were described and
compard with normal metalic phase and superconducting state
without vortices. Results were found as concerning the
influence of neighbouring planes on the vortex state in a given
plane and new exact solutions of the coupled Lagrange-Euler
sine-Gordon equations were found, their properties and free energy
were compared. Three-dimensional
superconducting state with a vortex in every plane with
cores localized in the same point
on a line perpendicular to planes
is a stable state in zero external magnetic field for
the layered materials. The same holds for a lattice of vortices
case. There exists a phase transition at the temperature
$T_{c}^{v}$ between the normal phase and
the vortex superconducting state which is of the first order.
Depending on the parameters of the material the transition
temperature $T_{c}^{v}$ may be higher than in the case of the
second order phase transition from normal phase to the
homogeneous (conventional) superconducting
phase without vortices.
The transition temperature $T_{c}^{v}$ is zero for zero
topological charge of vortices. Note that the transition is in
zero magnetic field. The first order phase transition has its
properties. It shows
usually  overheating and overcooling effects. Moreover
nucleation of the
superconducting phase in the normal phase occurs at temperatures
higher than the transition temperature $T_{c}^{v}$. And
vice-versa. Nucleation of the
normal phase in the superconducting phase occurs at temperatures
lower than the transition temperature $T_{c}^{v}$.
Thus the onset of the vortex-like excitations above
the transition temperature $T_{c}^{v}$ occurs in our theory due
to presence of nucleation of the superconducting phase with
vortices. The onset of the vortex-like Nerst signal above
the transition temperature $T_{c}^{v}$ in experiments in
LSCO, YBCO and in other
similar high-temperature materials which we discussed in the
Introduction may be explained by our theory. Vortex-like
excitations in $ La_{2-x}Sr_{x}CuO_{4} $ at
temperatures significantly above the critical temperature were
found in the Nerst effect signal \cite{XOWKU}. Such excitations
may correspond to formation of nuclei of the
vortex-superconducting phase above the transition temperature.
In overdoped $ La_{2-x}Sr_{x}CuO_{4} $ (LSCO) the upper
critical field $ H_{c2} $ does not end at $ T_{c0} $ but
at a much higher temperature. The upper critical field $ H_{c2} $
would correspond in our theory to a magnetic field at which pairs
in nuclei are destroyed, however firstly
vortex-antivortex dipols and prelattices should be destroyed.
Nerst measurements in $ YBa_{2}Cu_{3}O_{4} $ (YBCO) / and in
$ Bi_{2}Sr_{2-y}La_{y}CuO_{6} $ /, \cite{WXKUOAO},
and in LSCO, in fields up to 33 T show the existence
of vortex-like excitations high above $ T_{c0} $.
For these materials the same as above holds as concerning nuclei
of superconducting vortex phase and vortices.
The high field phase diagram of cuprates derived from the Nernst effect
implies that $ T_{c0} $ corresponds to a loss in phase rigidity
rather than a vanishing of the pairing amplitude,
\cite{WOXKUBLH}. Phase rigidity should be present for the phase
as the macroscopic order parameter. In nuclei this macroscopic
character is preserved, but nuclei have smalller volume than a
bulk superconducting phase. Thus there is a loss in phase
rigidity in the sense that nuclei are not correlated, macroscopic
phase in nuclei is not correlated too.
In zero magnetic field there is equal number of
vortices with plus and minus topological charge.
Vortex-antivortex pairs or parts of a vortex lattice are
strengthening presence of pairing amplitude due to selfconsistent
process described above. In nonzero field
there is nonequal number of vortices with plus and minus
topological charge.
Preformed pairs as superconductor fluctuations may exist in Bose
condensation of localised Cooper pairs with short coherence
length, \cite{YJU}, \cite{ASA} and \cite{MR}. Pairing
correlations without phase coherence are described in \cite{GB}
and \cite{EK}. Our theory is different from just mentioned
theories.

Some of recent reviews of the
high-temperature superconductors properties can be
found in \cite{PWA}, \cite{BV}, \cite{OM}, \cite{S}, \cite{TS}
and \cite{B1}. The experimental spectroscopic data are more
consistent. It is not quite clear what is the nature of
connection between antiferromagnetism and superconductivity.
Antiferromagnetic phase is the best understood part of the phase
diagram. ARPES methods lead to better understanding of the
quasiparticle dispersion \cite{RF}. Inelastic X-ray Raman scattering
\cite{MZH} give evidence for broken particle-hole symmetry.
There is present an incommensurate phase in antiferromagnetic
insulator phase \cite{CR}. The corresponding Hamiltonian contains
nearest neighbour, next-nearest neighbour magnetic interactions and the
interaction of spins of the fourth order. Magnetic Raman
scattering \cite{PES} shows that $A_{2}$ is very weak in $Gd_{2}
CuO_{4} $. In the phase diagram there is a pseudogap region
(phase) in YBCO, LSCO and BSCCO (2212). DC resistivity \cite{WT}
in BSCCO shows contribution of the pseudo-gap to resistivity
at high temperatures (180K - 200K). Pseudogap is seen
also in ARPES and tunneling experiments. While there are several
possible interpretations of the pseudo-gap region presence, like
stripes, antiferromagnetic fluctuations, spin-charge separation
and proximity to the QC point, the preformed pairs in
superconducting fluctuations above the critical temperature
is also one of such possibilities  \cite{B1}. These contributions may be
due to contribution of vortex-like excitations which we described
theoretically above.
Aharonov-Bohm effect is present in YBCO at 30K above the
critical temperature. Oscillations with flux period
h/2e are seen \cite{KK} . This may correspond to the
presence of the vortex-like excitations (vortex-antivortex pairs,
etc.) above the transition temperature. Phase coherence in BSCCO
was studied. Phase stiffness as a function of temperature at
frequencies of 100 GHz and 600 GHz was determined \cite{CJ}.
See also \cite{AJM}.
The Kosterlitz-Thouless superconducting transition is predicted
when the stiffness intersect this line increasing temperature.
It is interesting that high-frequency stiffness is non-vanishing
above the transition (short-length scales), and it is vanishing
at low frequencies (long-length scales). We have found \cite{OHRS}
that the inter-vortex distance (vortex lattice constant) for the
vortex state is increasing with temperature T and decreasing with
temperature T depending whether the temperature T is above or
below the critical transition temperature, respectively.
Fluctuations of the
vortex state above the critical transition temperature are bubbles
which may contain pairs or remnants of the vortex state. The
decrease of the inter-vortex distance with increasing temperature
in our theory is consistent with these phase stiffness measurements.
High-frequency superfluid stiffness is non-vanishing
(short-length scales) which would correspond to smaller
inter-vortex distance in our theory. Note that
we used $ \phi_{n}(\bf r) $, a change of the order parameter
phase around a vortex, for description of the system.
The constant amplitude approximation $ \mid
\Psi \mid^{2} = 1$ was considered, then we considered the
constant amplitude approximation in which the density of
superconducting pairs is temperature dependent. The constant
amplitude approximation is quite a good approximation due to the
fact that phase coherence and pairing amplitude are
present above the critical temperature in experiments:
the high field phase diagram
of cuprates implies that $ T_{c0} $ corresponds to a loss
in macrosocpic (bulk) phase rigidity
rather than a vanishing of the pairing amplitude,
\cite{WOXKUBLH}.

The inter-vortex distance is decreasing above this
temperature in our theory as it is observed in experiments with
phase stiffness, see above. The properties of $ \phi_{n}(\bf r) $,
 a change of the order parameter phase around a vortex, are more
 important than the changes of the density of pairs for
 description of the vortex state (phase), and thus
 changes of the amplitude of the order parameter are less
 important.

In this paper we reviewed our results on
vortex configurations in two-dimensional sine-Gordon
systems in connection with high-temperature superconductors.
In \cite{OHR} and \cite{OHRS}.
we discussed the vortex states using exact solutions
of these coupled Lagrange-Euler sine-Gordon equations,
their properties and their free energy was compared with the
normal phase state and the homogeneous (conventional) superconducting state.
Stable vortex configurations in zero external magnetic field for
the layered materials exist, the same holds for a lattice of vortices
case. The first order phase transition overheating and overcooling
effects are present in the system. Nucleation of the
superconducting phase in the normal phase, which occurs at temperatures
higher than the transition temperature $T_{c}^{v}$, then occurs.
The onset of the vortex-like excitations above
the transition temperature $T_{c}^{v}$, which occurs in our
theory, may correspond to the onset of the vortex-like Nerst signal above
the transition temperature $T_{c}^{v}$ in LSCO, YBCO and in other
similar high-temperature materials. Thus the vortex-like
excitations not well understood in experiments may be compared
with properties of vortices from our theory. It was found
experimentally  that the vortex-like excitations above
the transition temperature $T_{c}^{v}$ in LSCO, YBCO and in other
similar high-temperature materials continuously evolve as
concerning their properties to vortex states below
the transition temperature $T_{c}^{v}$. This fact may be
explained within our theory. Thus our theory may explain some of
the experimentally  observed properties of high-temperature
superconductors described above. It will be interesting to study
and compare other properties of vortex-like excitations and
vortex states in high-temperature materials
and compare these properties with those described by anisotropic vortices.

\section*{Acknowledgements }

The author wishes to express his sincere thanks to the Abdus
Salam International Center for Theoretical Physics for the
invitation to present some of the results presented in this paper
at the 12th Workshop on Strongly Correlated Electron Systems,
2000 and at the Workshop Emergent Materials and Highly Correlated
Electrons, 2002. The papers contains some of the results presented
in the invited talk given at the Second International School
on Condensed Matter in Debrecen, September 2000, the author
especialy wishes to express his sincere thanks to
Zs. Gulaczi for his interest. The author would like to express
his thanks for invitation to present part of the results
presented in this paper on the Second International Workshop on Electron Correlations and Materials
Properties, Rhodos, Greece, 25th Jun - 29th Jun 2001 to
organizers, especially his sincere thanks are due to Dr. A. Gonis.
The paper was supported by the grant VEGA No. 1/0250/03,
Department of Theoretical Physics, Faculty of Mathematics,
Physics and Informatics, Comenius University, Bratislava.

\end{document}